\documentclass{elsart}

\usepackage{epsfig}
\date{today}

\begin{document}

\begin{frontmatter}

\title{Phonons Softening in Tip-Stretched Monatomic Nanowires}

\author[SISSA]{Fabien Picaud\thanksref{fellow}} ,
\author[SISSA,INFM]{Andrea Dal Corso}
\author[SISSA,INFM,ICTP]{and Erio Tosatti}
\address[SISSA]{SISSA Via Beirut 2/4, 34014 Trieste (Italy)}
\address[INFM]{INFM, Unit\`a di Trieste Via Beirut 2/4, 34014 Trieste
(Italy)}
\address[ICTP]{ICTP, Strada Costiera 11, 34014 Trieste (Italy)}
\thanks[fellow]{Fellow of TMR FULPROP, EU Contract ERBFMRXCT970155.}

\maketitle

\begin{abstract}
It has been shown in recent experiments that electronic transport through
a gold monatomic nanowire is dissipative above a threshold voltage due to
excitation of phonons via the electron-phonon interaction.
We address that data by computing, via density functional
theory, the zone boundary longitudinal phonon frequency of a perfect
monatomic nanowire during its mechanical elongation. The theoretical
frequency that we find for an ideally strained nanowire is not compatible
with
experiment if a uniformly distributed stretch is assumed. With the help
of a semi-empirical Au-Au potential, we model the realistic nanowire
stretching as exerted by two tips. In this model we see that strain
tends to concentrate in the junctions, so that the mean strain of the
nanowire is roughly one half of the ideal value. With this reduced strain,
the calculated phonon softening is in much better agreement with
experiment.

Keywords : Nanowires, Gold, Conductance, Dissipation,
Phonons, Softening.

\end{abstract}
\end{frontmatter}

\section{Introduction}

Quantized electronic transport through metallic gold nanowires has been
extensively studied experimentally~\cite{YANSON}. Recent transmission
electron microscopy and break junction data showed that unit conductance
(in unit of $G_0 = 2e^2/h$) corresponds to tip-suspended
wires segments consisting of a single
monatomic strand~\cite{RODRIGUES,OHNISHI,SMIT}. At sufficiently low
voltage,
the conductance of these wires is ballistic. However, a recent
experiment \cite{AGRAIT} showed that above a voltage threshold of
about 15 meV a monatomic gold wire, seven atoms long, displays a drop of
about 1 \% in the differential conductance below the ballistic value of
one. The
symmetric drop of conductance with respect to the voltage sign and the
value of the voltage threshold of around 15 meV suggested the effect
to be due to inelastic scattering of electrons by
longitudinal wire vibrations. Under that hypothesis, the voltage
threshold yields precisely the frequency of the most important
nanowire phonons. That voltage was shown experimentally to drop
indicating a phonon softening with increasing stretching, and that needs
to be quantitatively explained.
Here we present calculations, based on density
functional theory, of the vibrational frequencies of a monatomic
gold wire and study their dependence on the wire stretching. Since
simple arguments suggest that the phonon responsible for the effect is the
longitudinal acoustic phonon at the zone boundary $\pi/a$ ($a$ is the
gold-gold distance),
we concentrate on the frequency of this vibration.

We show that the
dependence of the longitudinal frequency on wire strain is not
compatible with experiment if a uniformly distributed nominal strain is
assumed. In order to analyze the role of possible non-uniform strain
configurations,
we modeled the mechanics of a realistic nanowire in contact with two tips
with a semi-empirical gold potential. We found that strain was
non-uniformly distributed, the tip-wire junctions absorbing a large
fraction, and reducing correspondingly the true nanowire strain.
With this reduction the agreement of experimental and theoretical
stretch-dependent phonon frequency was very much improved.

\section{Phonon Calculations for a Stretched Nanowire}

We considered an infinite monatomic nanowire
and calculated its phonon frequencies as a function of the wire
strain, that is for increasing interatomic spacing $a$.  The model is
of course very crude since it does not include at all the wire-tip
junctions; but as a simple approximation it allows us to address
first of all the reduced dimensionality effect on the force constants among
gold atoms. The nanowire total energy and electronic structure were
determined within standard Density Functional Theory (DFT) in the local
density approximation. The phonon frequencies were calculated within
density functional perturbation theory, using the PWSCF and PHONON
codes~\cite{PWSCF} with the gold pseudopotential of Ref. \cite{TOSATTI}.
The gold wire was simulated by using a tetragonal periodically repeated
supercell. The wire was oriented along the $z$ axis
and repeated periodically along $x$ and $y$ with a wire-wire distance
($10.58$ \AA) large enough to avoid unphysical interactions among
replicas.
The energy cutoffs of the plane wave basis was $28$ Ry for the
wavefunctions and $252$ Ry for the charge density.
We used $40$ uniformly distributed {\bf k} points to
integrate the one dimensional Brillouin zone and a
broadening technique~\cite{METHFESSEL} with a smearing parameter
of $0.01$ Ry to deal with the presence of a Fermi level.
These parameters were found to be sufficient for a good convergence
of the results.
Based on these first principles calculations, we also generated
parameters for an empirical effective potential between gold atoms.
In a tight binding scheme, and in the second moment approximation (SMA),
the effective potential between Au atoms separated by $r_{ij}$ may be
written as
\cite{GAMBA}:
\begin{equation}
E_i=\lambda \sum_je^{-p\left( \frac{r_{ij}}{r_0}-1\right)
}-\epsilon \left( \sum_je^{-2q\left( \frac{r_{ij}}{r_0}-1\right)
}\right) ^\alpha \textit{ with} \ r_{ij}<r_c  \label{EQUA1}
\end{equation}
The parameters entering in the above expression can be obtained by fitting
the first-principles energies and related bulk properties of gold. We
obtain in this way $\lambda=0.4086$,
$p=8.5624$, $\epsilon=1.6332$, $q=3.6586$, $\alpha=0.6666$, $r_0=2.88$
\AA.

\section{Results and discussion}
In our first step, we calculated with DFT the equilibrium
properties of the infinite monatomic wire (Fig. 1).  The stable chain
under zero tension (minimum of the total energy) has a nearest
neighbour distance $a_{0}$=2.49 \AA\ in agreement with other calculations
\cite{TORRES,OKAMOTO,SANCHEZ}.  Neglecting electron reflection at the
junctions (justified for a very wide s-band) this ideal wire has a
ballistic conductance of one due to a single band
crossing the Fermi level at $\pi$/$2 a_{0}$. The important resistive
electron-phonon coupling is electron backscattering by the
zone boundary longitudinal phonon at $q_z = \pi /a_{0}$.
At the strain-free equilibrium spacing $a_{0}$,
the calculated phonon frequencies are $240$ cm$^{-1}$ and $ i52$ cm$^{-1}$
for the longitudinal and the two degenerate transverse modes, respectively.
As a reference, with these pseudopotentials, the stretch frequency of
the Au dimer is $191$ cm$^{-1}$.
The imaginary frequency of transverse modes indicates that an
infinite wire is unstable under zero tension, for example against
a zigzag distortion, as discussed in Ref.\cite{SANCHEZ}. This
in no way implies that the real wire segment hanging between the two
tips will be distorted, since the equilibrium geometry will
be strongly influenced by the tips. In particular, grand canonical
tip-wire equilibrium requires a positive string tension \cite{TOSATTI},
and an increase of $a$. We find that the frequency of the transverse
modes becomes positive for $a$=2.75 \AA.

Returning to the evolution of the frequency of the longitudinal phonon
under wire stretching, we calculated that by increasing progressively
the distance between atoms in the chain. We show in Fig. 2
the evolution of the vibrational frequency $\omega_{X}$ at the $X$ point
as a function of the distance $a$. During stretching the bond softens
and $\omega_{X}$ decreases almost linearly as $a$ increases.
At $a$=2.85 \AA, $\omega_{X}$ becomes negative and above this
distance the infinite wire is unstable, for example against dimerization.
For comparison, we also plotted in Fig. 2 the
frequencies deduced from the experiment of Ref. \cite{AGRAIT} (long dashed
line).
In order to decide the position of these points we made two assumptions.
First we converted the measured elongation, $0.75$ \AA, into an
elongation of the gold-gold bonds, assuming that the experimental wire is
uniformly stretched and dividing the total elongation on each bond of
the $7$ atom wire. Next we had to decide what was the starting gold-gold
distance
in the experimental curve. Since the initial wire tension is unknown,
we assumed the highest measured longitudinal
frequency to correspond to a bond length equal to $2.68$ \AA.
In this way the experimental and theoretical frequencies of the first
point coincide. The slopes of the two curves however differ by a factor of
almost $2$, theory decreasing faster than experiment.

This discrepancy between the calculated dependence of the
longitudinal frequency on the gold-gold distance and the measured one
is too large to be attributed to computational inaccuracy, and
rather suggests that we made a wrong assumption somewhere.
Our main doubt eventually converged on the assumption of
uniform strain. On the one hand the gold-gold bond could be expected
to be stronger in the monatomic nanowire than in the tip, as
indicated by shorter distances. On the other hand the
dependence of the experimental phonon frequency with strain
would be in better agreement with the theoretical values if one
were to assume a elongation of $0.04$ \AA\ instead of nearly $0.1$ \AA\
(7 Au atoms and a total elongation of $0.75$ \AA).

A full ab-initio attack of this problem is desirable but seems well
beyond the scope of our work. Thus,
we investigated further this aspect by simulating with our classical
SMA effective potential a realistic 7-atom
chain in contact with two tips, where all atoms in the chain and in
the tips are allowed to move without any geometrical constraints.
It is important to note that the mechanical properties of the gold wire
calculated by DFT and SMA differ by very little, essentially only a small
shift of the equilibrium spacing (about $0.05$ \AA)
and the cohesive energies differs by only $0.23$ eV (see Fig. 1).
The gold chain is thus obtained by optimizing the total energy of the
atoms in the system while the distance of the two tips is increased. Each step
corresponds to an elongation of $0.02$ \AA. At each step, the equilibrium
configuration attained is only guaranteed to be a local minimum of
the potential energy surface, which in turn depends on the starting
configuration. Nevertheless, the strain evolution obtained
appeared very reasonable, providing a vivid picture
of a possible inhomogeneous strain in the tip-suspended nanowire.

Fig. 3 displays typical snapshots during the formation and the
stretching of a wire. At the beginning of the stretching (Fig.
3a), the two tips are stuck together. Upon stretching a twisted chain
appears
while one atom is extracted. Subsequently a chain of 8 atoms is
formed by successive incorporation of atoms from the tips. At this
length, the simulated nanowire broke.  Incorporation of each new
atom in the chain was observed after a elongation of
nearly 1.5 \AA\ and involved a sudden downward jump in bond lengths.
This crude model shows that the wire of given length can only be
stretched elastically over a limited distance which compares well
with the 1 \AA\ in the experiments \cite{AGRAIT}.
The calculated change of nanowire interatomic distance for a total
elongation
tip-tip distance of $0.75$ \AA\ and a chain of 7 atoms was  $0.03$ \AA,
instead of nearly
$0.1$ \AA\ as expected the perfect chain. Using a
nanowire strain corrected by this scaling factor, the experimental phonon
softening against corrected strain (dot-dashed line in figure 2) is now
in much better agreement with theory.

\section{Conclusion}
We calculated the evolution with stretching of the longitudinal
phonon frequency of a monatomic gold wire, motivated by recent observations
indicating a softening. The softening obtained via density functional
theory in a monatomic wire is indeed large, in fact too large. It is
compatible
with the experimental data only if a non uniform distribution of the
strain is introduced, the largest strain being taken up by the
tips and tip-wire junctions, and only 40\% of the strain in the chain.

\ack{\vspace{-0.4cm} This project is sponsored by MIUR COFIN2001,
INFM/F, INFM/G, Iniziativa Trasversale Calcolo Parallelo, and by
EU through TMR FULPROP, Contract ERBFMRXCT970155.}

\newpage


\begin{figure}
    \begin{center}
        \includegraphics[width=10cm,height=10cm]{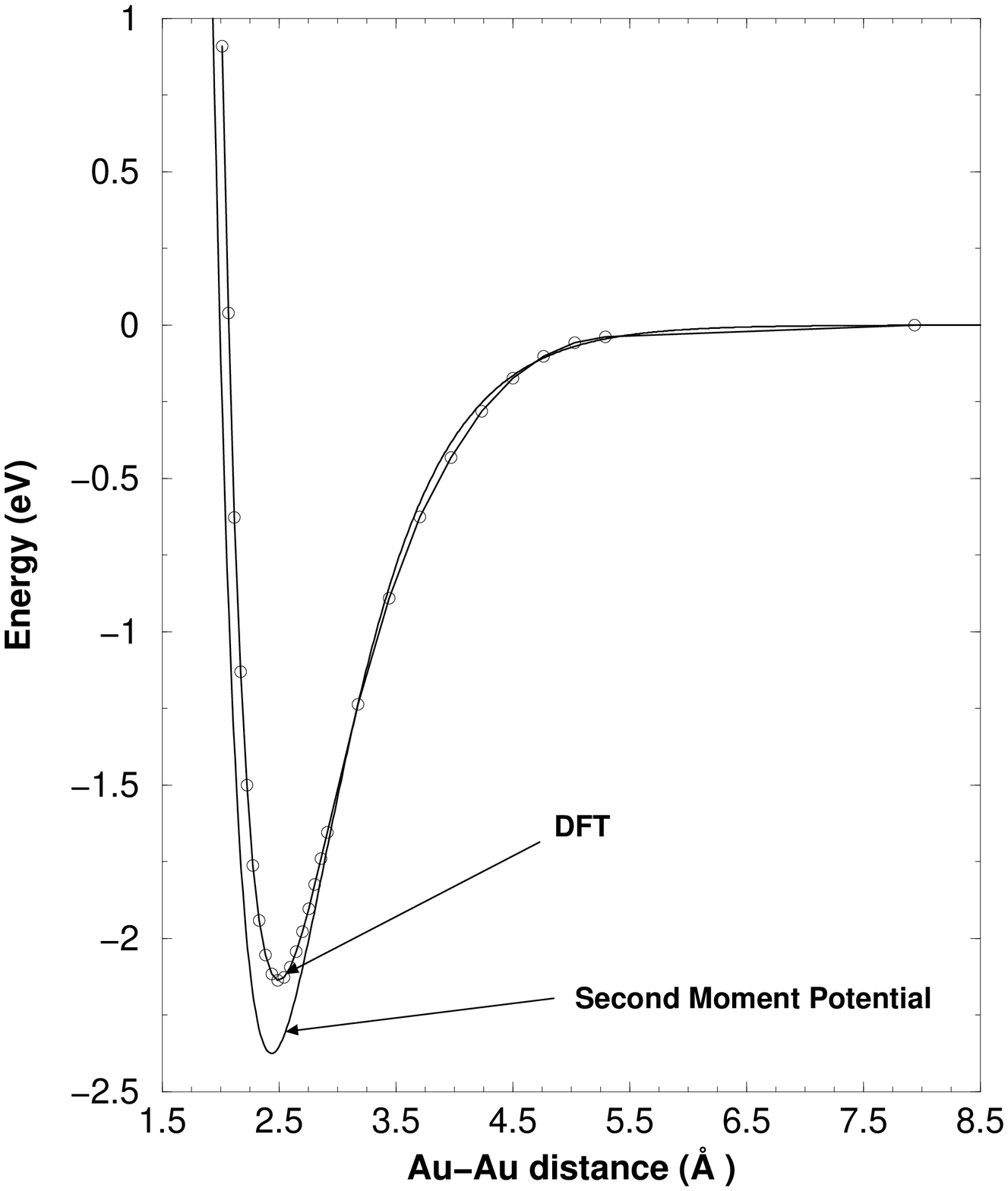}
    \end{center}
\end{figure}

Fig. 1: Circles and continous line: total energy of the infinite monatomic
gold wire obtained by DFT-LDA as a function of the Au-Au distance.
Dashed line: total energy of the infinite monatomic gold wire
calculated with the semi-empirical potential\cite{GAMBA}.

\newpage
\begin{figure}
    \begin{center}
        \includegraphics[width=10cm,height=10cm]{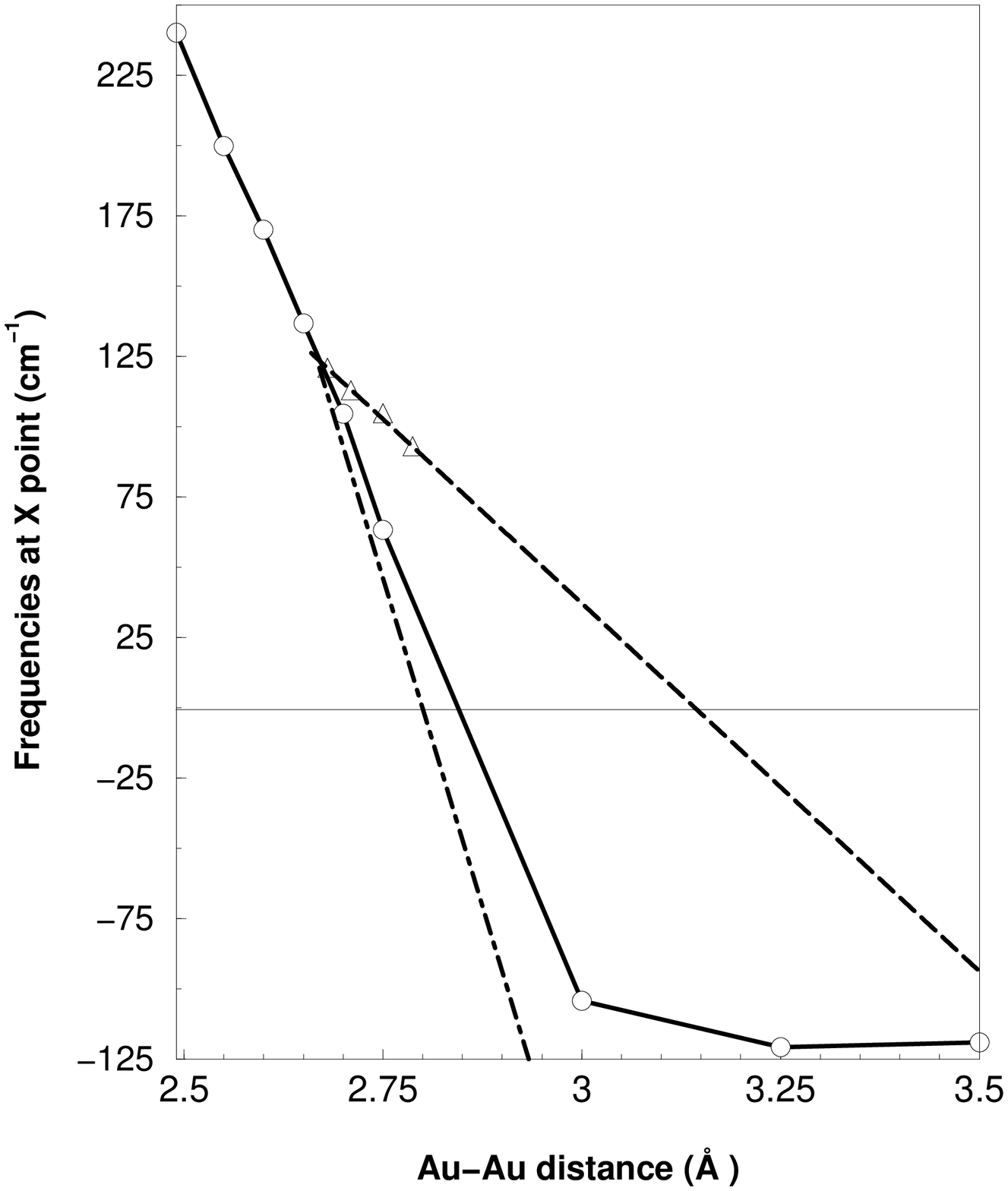}
    \end{center}
\end{figure}
Fig. 2: Frequencies of the longitudinal phonon as a function of the
gold-gold distance. Continuous line and circles correspond
to theoretical results. Triangles are the experimental data in the first
linear part of Fig. 3d of Ref.\cite{AGRAIT}. The dashed line is a linear
interpolation of experimental data. The
dot-dashed line represents the experimental frequencies replotted
for a stretching rescaled by a factor 0.3 (see text).

\newpage
\begin{figure}
    \begin{center}
        \includegraphics[width=15cm,height=15cm]{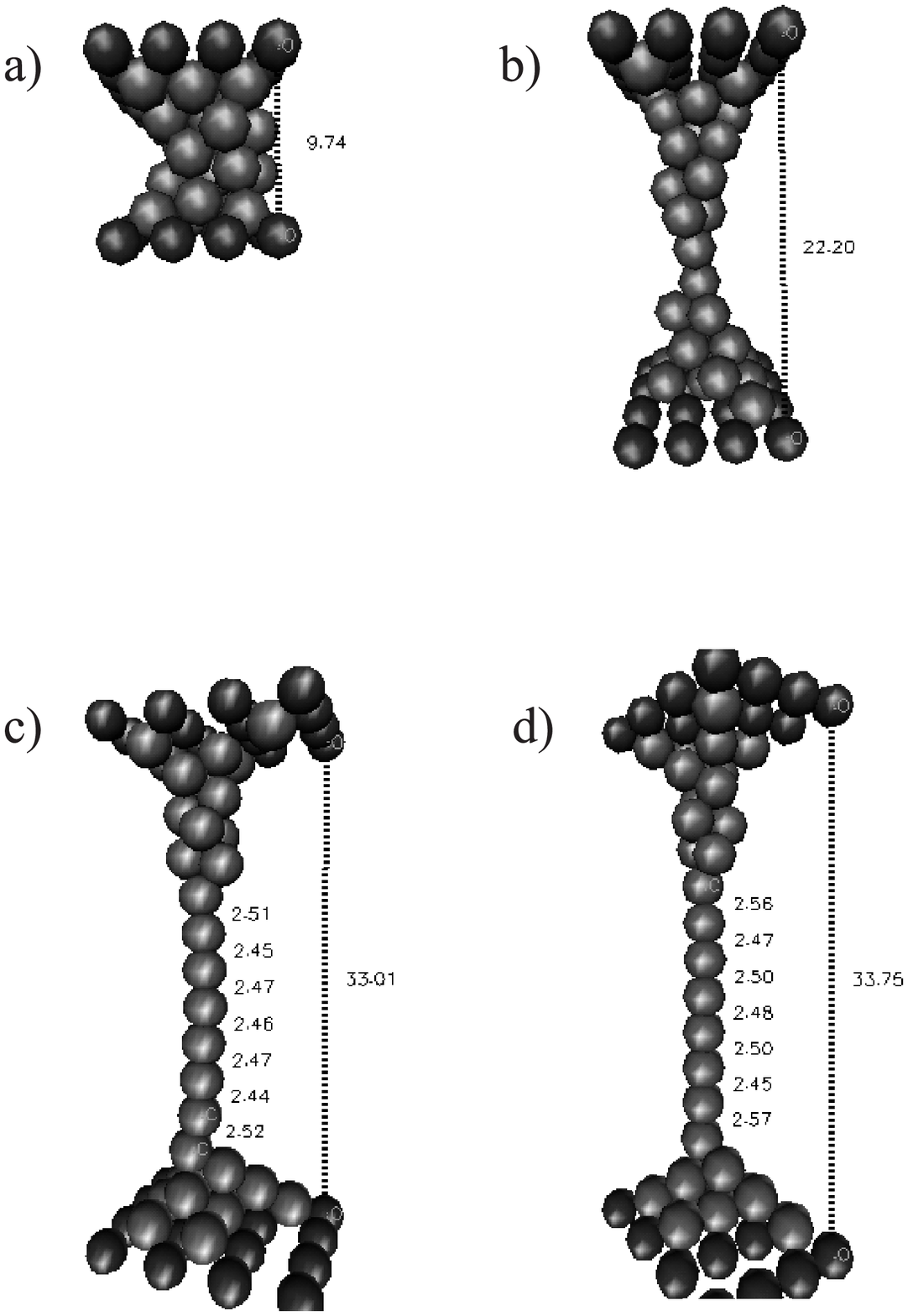}
    \end{center}
\end{figure}

Fig. 3: Mechanical model. The total energy of the system is
minimized during the stretching of the two tips. a) Starting
configuration of the simulation. The two tips are in the nearest
position. b) One atom is extracted . c) Strained nanowire 7 atoms long. The
lengths of each bond are indicated on the snapshot. d) After
elongation of $0.75$ \AA\ of the 7 atoms wire, one supplementary
atom is extracted. At this moment, the evolution of the bonds
shows a mean deviation of $0.03$ \AA\. The wire broke in this
simulation before extraction of the 9th atom.

\end{document}